\documentclass{appolb}
\usepackage{graphicx}
\usepackage[english]{babel}
\usepackage[utf8]{inputenc}
\usepackage{graphicx,xcolor}
\usepackage{epsf}
\usepackage{amssymb}
\usepackage{cite}
\usepackage{float}
\usepackage{hyperref}
\usepackage{booktabs}
\usepackage{siunitx}
\usepackage{hyperref}
\usepackage{placeins}
\usepackage{epstopdf}
\usepackage{listings} 
\newcommand{\beq}{\begin{equation}}
\newcommand{\eeq}{\end{equation}}
\newcommand{\bea}{\begin{eqnarray}}
\newcommand{\eea}{\end{eqnarray}}

\begin{document}
\title{Inclusive hadron-jet production at the LHC %
\thanks{Presented by F.G~Celiberto at \bf{\textit{Diffraction and Low-x 2018}}.}%
}
\author{A.D.~Bolognino$^{1,2}$, \underline{F.G.~Celiberto}$^{3,4}$, D.Yu.~Ivanov$^{5,6}$, M.M.A.~Mohammed$^{1}$, A.~Papa$^{1,2}$
\address{
\centerline{${}^1$ {\sl Dipartimento di Fisica, Universit\`a della Calabria, I-87036 Arcavacata di Rende,  Cosenza, Italy}}
\vspace{0.1cm}
\centerline{${}^2$ \sl Istituto Nazionale di Fisica Nucleare, Gruppo collegato di
Cosenza,} \centerline{\sl I-87036 Arcavacata di Rende, Cosenza, Italy}
\vspace{0.1cm}
\centerline{${}^3$ {\sl Dipartimento di Fisica, Universit\`a degli Studi di Pavia, I-27100 Pavia, Italy}}
\vspace{0.1cm}
\centerline{${}^4$ \sl Istituto Nazionale di Fisica Nucleare, Sezione di Pavia, I-27100 Pavia, Italy}
\vspace{0.1cm}
\centerline{${}^5$ {\sl Sobolev Institute of Mathematics,
630090 Novosibirsk, Russia}}
\vspace{0.1cm}
\centerline{${}^6$ {\sl Novosibirsk State University, 630090 Novosibirsk, Russia}}
}}

\maketitle
\begin{abstract}
We suggest the inclusive detection at the LHC of a light charged hadron and of a jet widely separated in rapidity as a new probe channel for the study of the BFKL resummation. Predictions for cross section and azimuthal correlations, shaped on the CMS and CASTOR acceptances, are presented.
\end{abstract}
  
\section{Introduction}

Semi-hard reactions represent a challenging testfield for perturbative QCD in the high-energy limit. Here, the fixed-order description undershoots the effect of large energy logarithmic contributions, which balance the smallness of the QCD coupling constant and hence must be resummed to all orders. The Balitsky-Fadin-Kuraev-Lipatov (BFKL)~\cite{BFKL} approach is the most powerful tool to resum to all orders these large logarithms both in the leading (LLA) and the next-to-leading (NLA) approximation. In the last years, a good number of processes have been proposed as candidate probes of the high-energy regime, namely: the inclusive hadroproduction of two jets with high $p_T$ and well separated in rapidity (better known as Mueller-Navelet process~\cite{Mueller:1986ey}), for which a richness of theoretical predictions have appeared so far~\cite{Colferai:2010wu,Angioni:2011wj,Caporale:2012ih,Ducloue:2013wmi,Ducloue:2013bva,Caporale:2013uva,Ducloue:2014koa,Caporale:2014gpa,Ducloue:2015jba,Caporale:2015uva,Celiberto:2015yba,Celiberto:2016ygs,Chachamis:2015crx,Caporale:2018qnm}, the inclusive detection of two identified, light charged hadrons~\cite{Ivanov:2012iv,Celiberto:2016hae,Celiberto:2017ptm}, and more recently, multi-jet hadroproduction~\cite{Caporale:2015vya,Caporale:2015int,Caporale:2016soq,Caporale:2016xku,Celiberto:2016vhn,Caporale:2016pqe,Caporale:2016zkc} and heavy-quark pair photoproduction~\cite{Celiberto:2017nyx}.
In this work a novel semi-hard reaction, {\it i.e.} the concurrent detection of a light charged hadron, $\pi^{\pm}, K^{\pm}, p \left(\bar p\right)$, and a
jet, both with high $p_T$ and separated by a large rapidity interval, is proposed and investigated in the NLA BFKL approach. Although being, {\it de facto}, a hybridization of two already studied processes, this channel presents some own characteristics which can make it worthy of consideration in future analyses at the LHC. First, the tag of two different kinds of final-state object leads to a natural asymmetric configuration in $p_T$, allowing us to suppress the Born contribution and enhancing the discrepancy between BFKL and DGLAP, as pointed out in~\cite{Caporale:2014gpa,Celiberto:2015yba,Celiberto:2015mpa}. Second, having just one final-state hadron identified, instead of two, should quench ``minimum-bias'' contaminations, thus facilitating the comparison with experimental analyses. Third, one can use this reaction to compare different sets of fragmentation functions (FFs) and jet algorithms, dealing with {\it linear} expressions in the corresponding functions. 

\section{Inclusive hadron-jet production at the LHC}

The process under investigation is~\footnote{This process has much in common with the inclusive $J/\Psi$-meson plus backward jet production, considered recently in~\cite{Boussarie:2017oae}.}
\begin{eqnarray}
\label{process}
{\rm proton}(p_1) + {\rm proton}(p_2) 
\to 
{\rm hadron}(k_H, y_H) + {\rm X} + {\rm jet}(k_J, y_J) \;,
\end{eqnarray}
when a light charged hadron, $\pi^{\pm}, K^{\pm}, p \left(\bar p\right)$, and a
jet, featuring large transverse momenta, $k_{H,J} \gg \Lambda_{\rm QCD}$, and well separated in rapidity, $Y \equiv y_H - y_J$, are produced together with an undetected hadronic system, X. 
The differential cross section of the process can be presented as
\begin{equation}
\frac{d\sigma}
{dy_Hdy_J\, d|\vec k_H| \, d|\vec k_J|d\phi_H d\phi_J}
=\frac{1}{(2\pi)^2}\left[{\cal C}_0+\sum_{n=1}^\infty  2\cos (n\phi )\,
{\cal C}_n\right]\, ,
\end{equation}
where $\phi \equiv \phi_H - \phi_J - \pi$, with $\phi_{H,J}$ the hadron/jet azimuthal angles, while ${\cal C}_0$ is the $\phi$-averaged cross section and the other coefficients ${\cal C}_n$ determine the azimuthal-angle distribution of the final state. 
In order to match realistic kinematic configurations used at the LHC, we integrate the coefficients
over the phase space for two detected objects and keep fixed the rapidity interval, $Y$, between the hadron and the jet: 
\begin{equation}
\label{Cn_int}
C_n= 
\int_{y^{\rm min}_H}^{y^{\rm max}_H}dy_H
\int_{y^{\rm min}_J}^{y^{\rm max}_J}dy_J\int_{k^{\rm min}_H}^{k^{\rm max}_H}dk_H
\int_{k^{\rm min}_J}^{{k^{\rm max}_J}}dk_J
\, \delta \left( y_H - y_J - Y \right)
\, {\cal C}_n 
%\left(y_H,y_J,k_H,k_J \right)
\, .
\end{equation}

\begin{figure}[H]

  \centering
  \includegraphics[scale=0.26,clip]{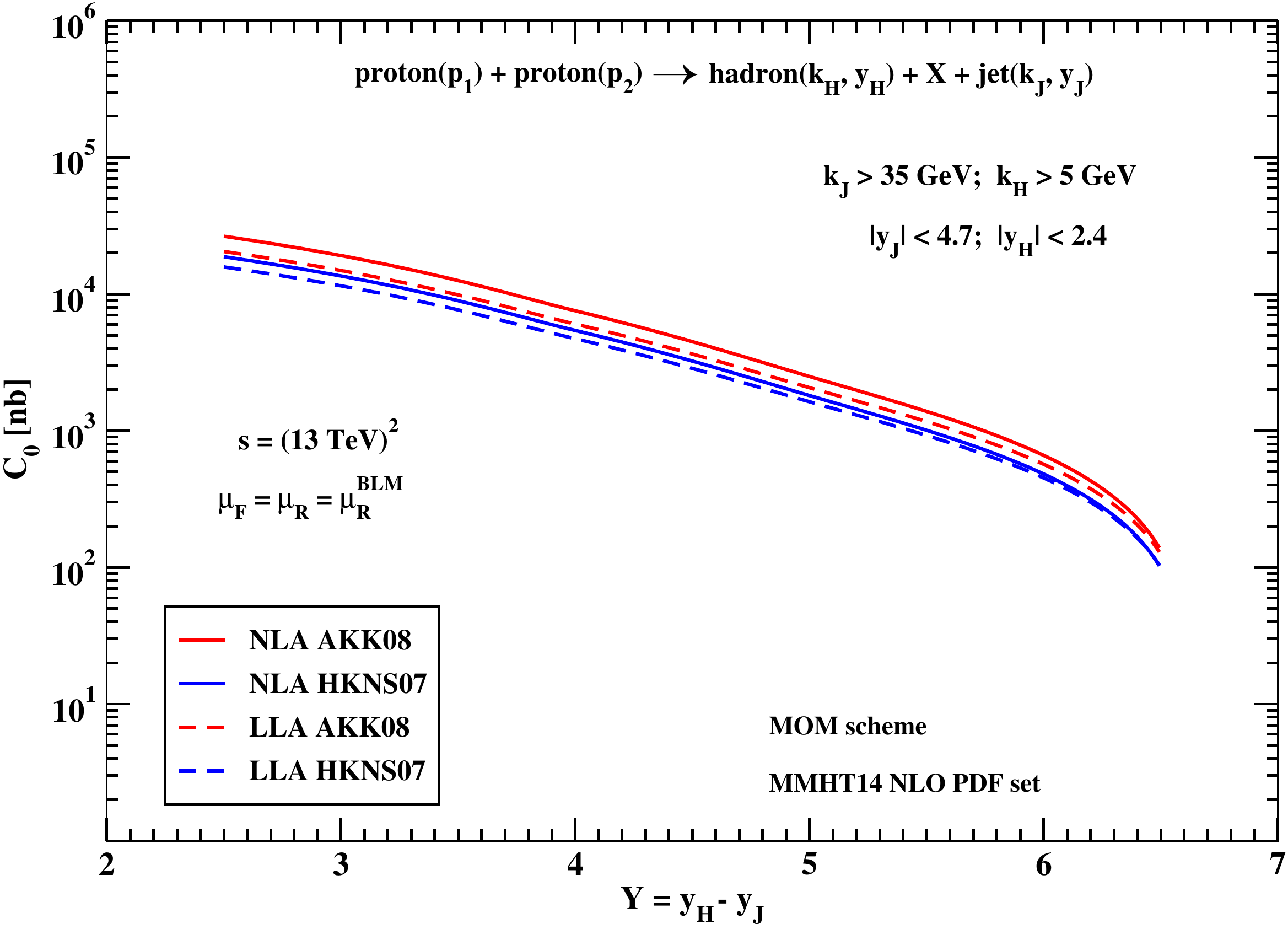}
  \includegraphics[scale=0.26,clip]{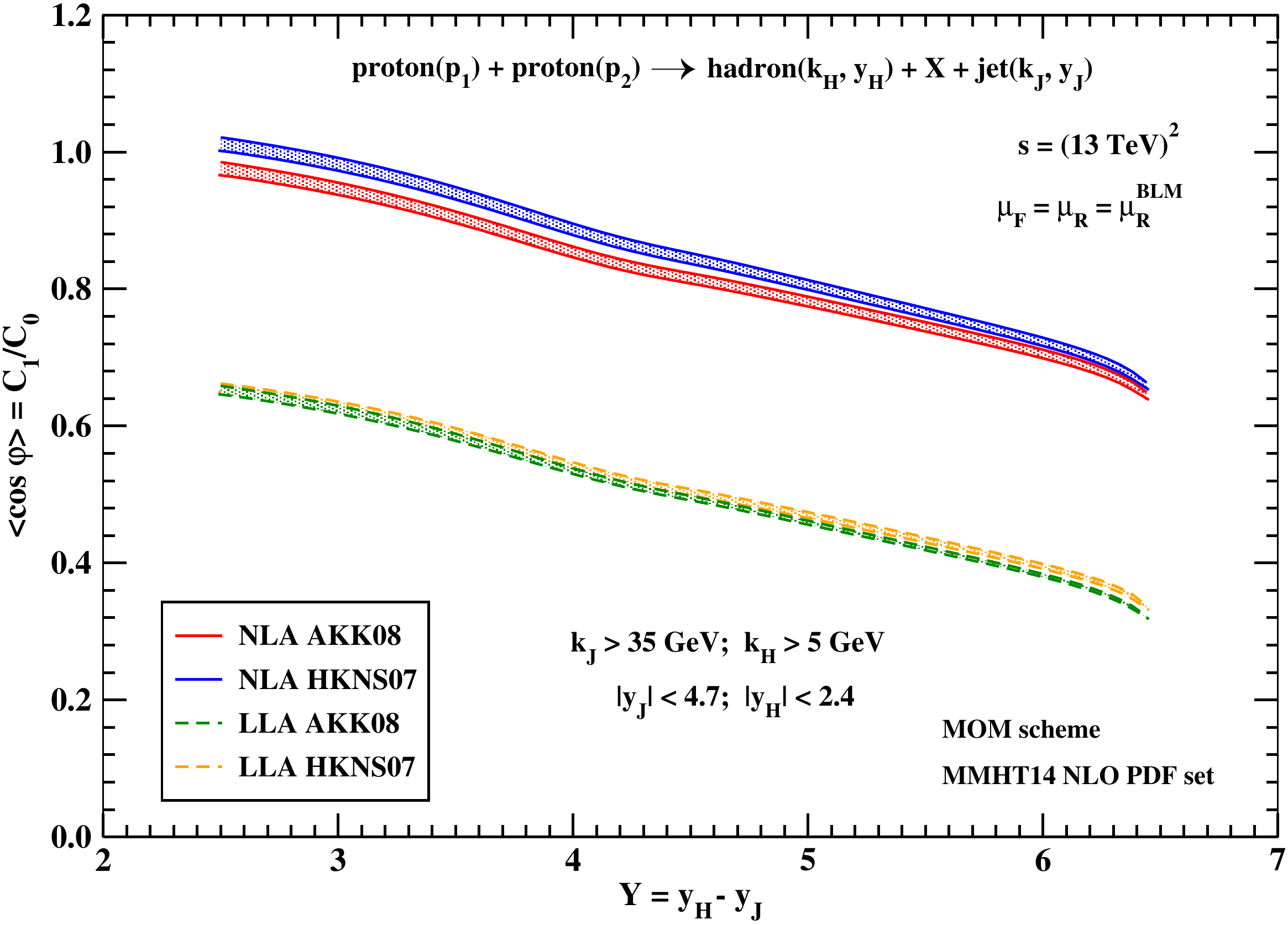}

  \includegraphics[scale=0.26,clip]{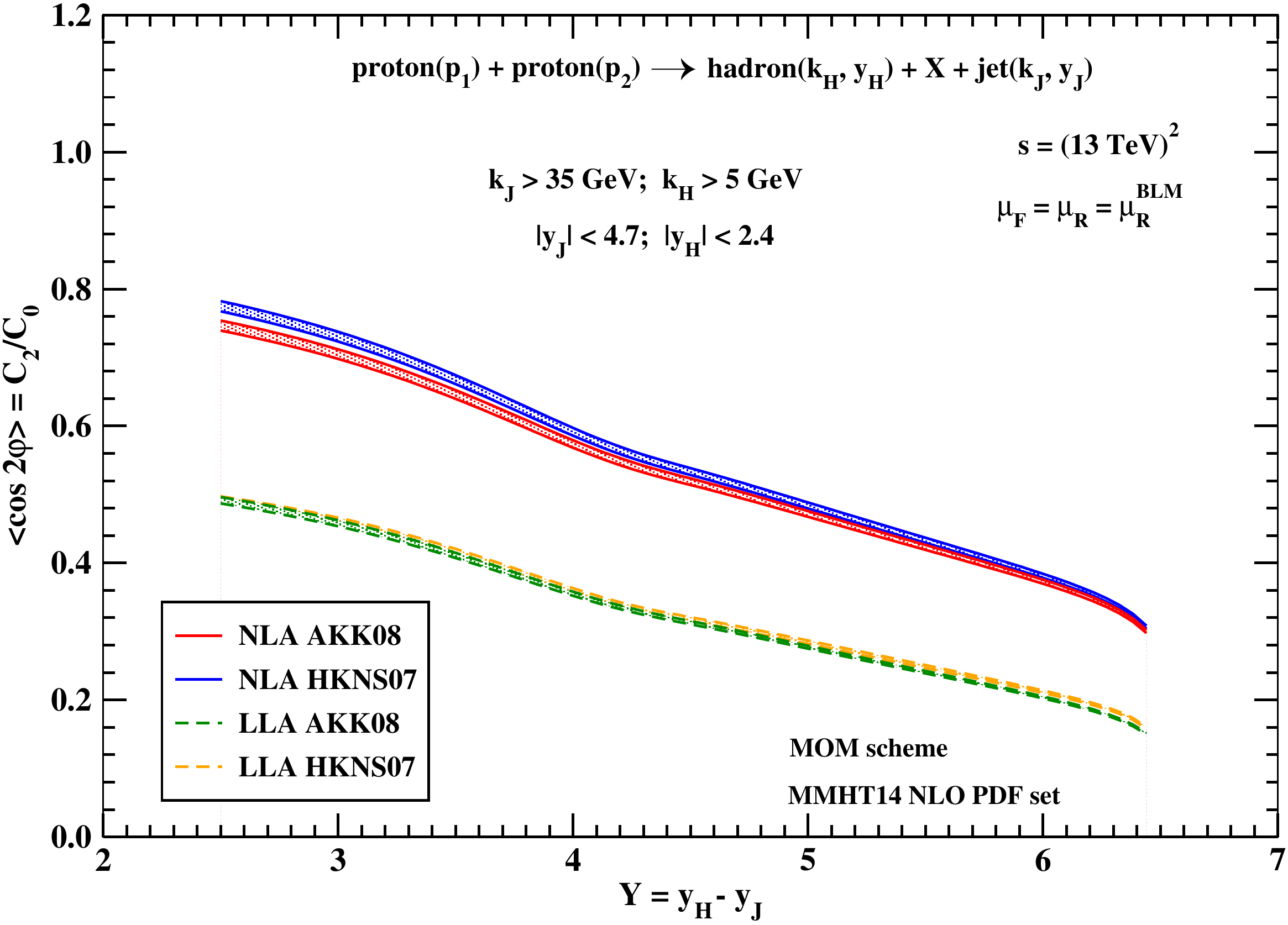}
  \includegraphics[scale=0.26,clip]{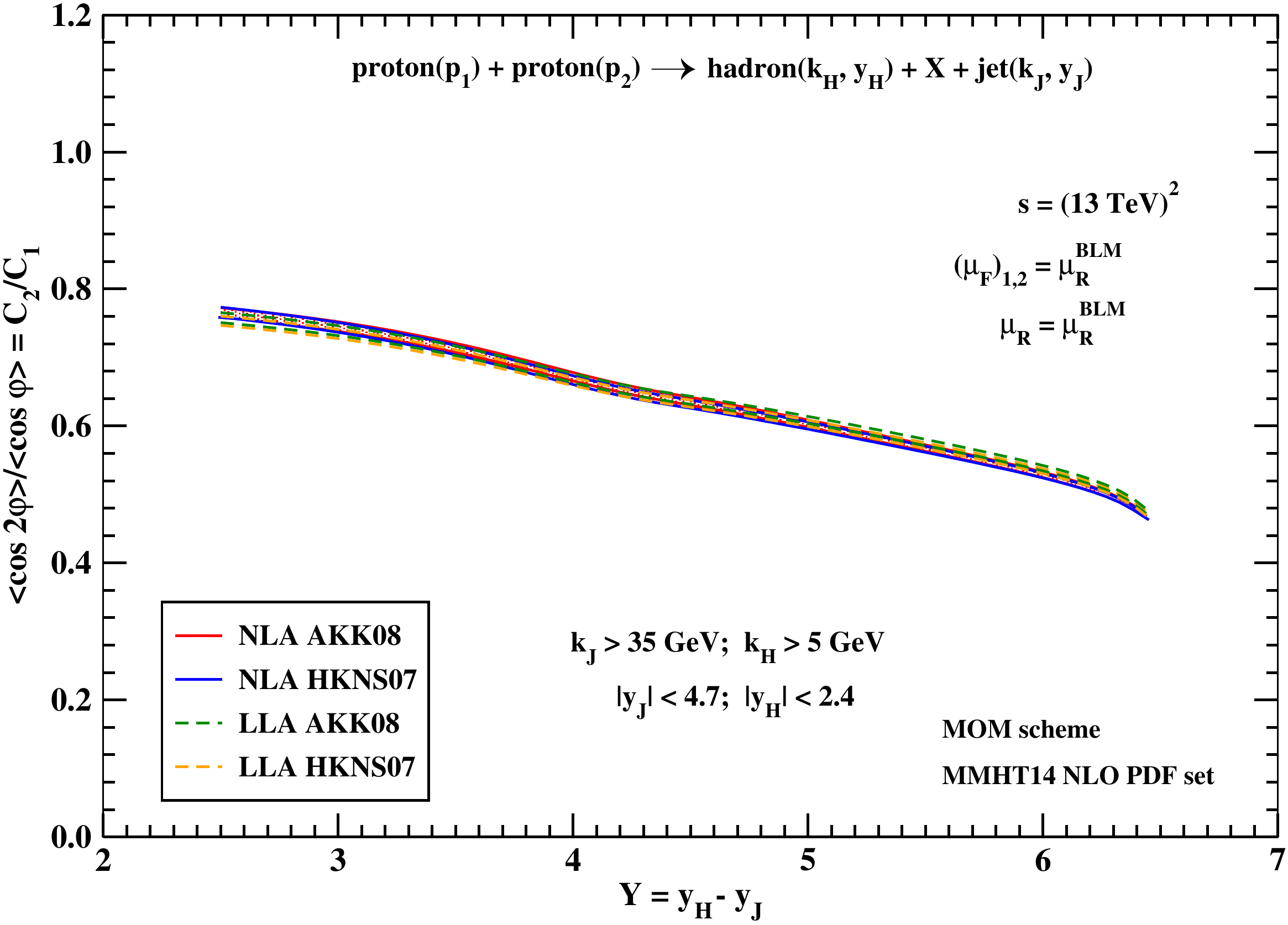}

  {\it a}) {\it CMS-jet} configuration.
  \vspace{0.5cm}

  \includegraphics[scale=0.26,clip]{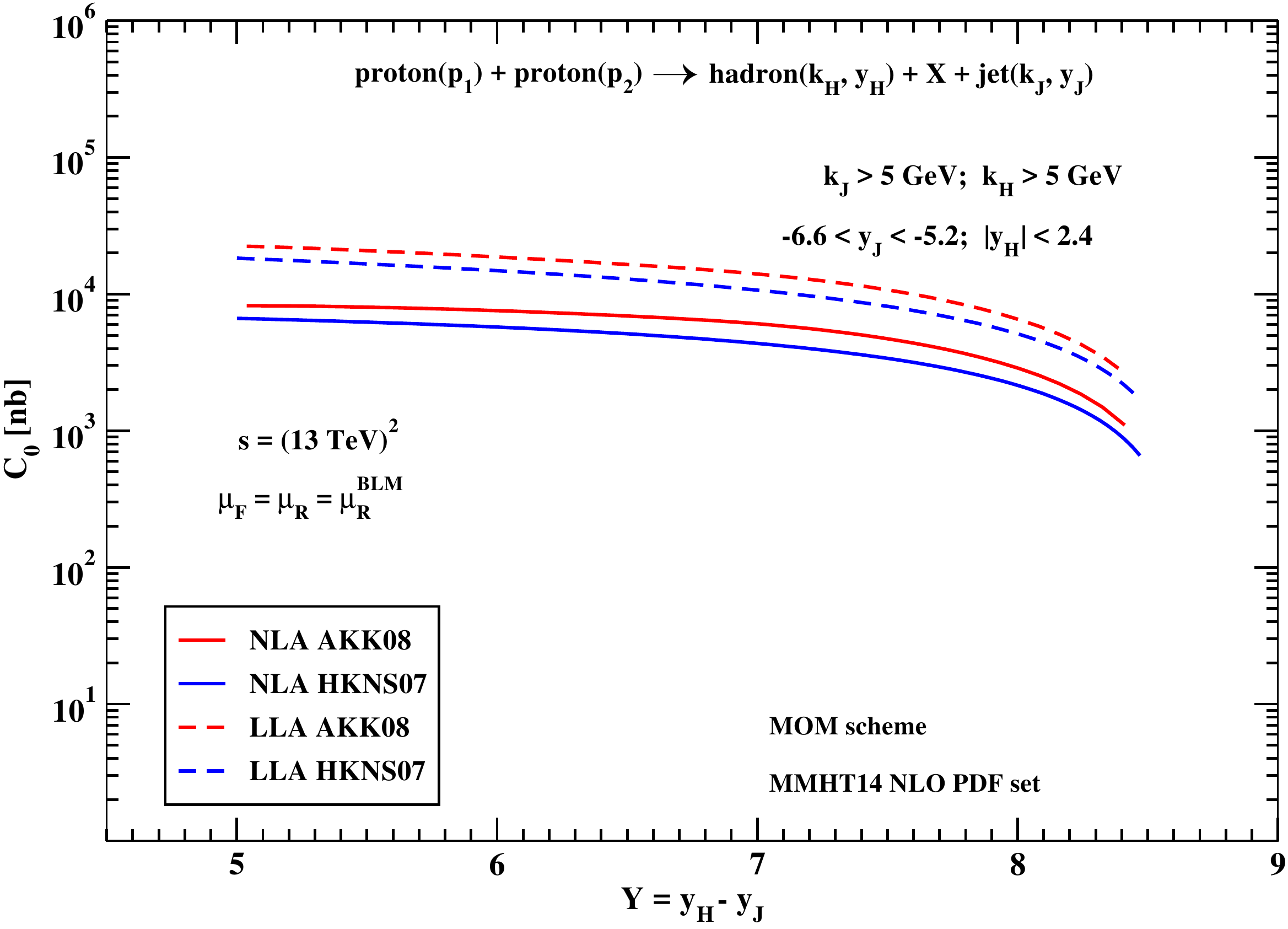}
  \includegraphics[scale=0.26,clip]{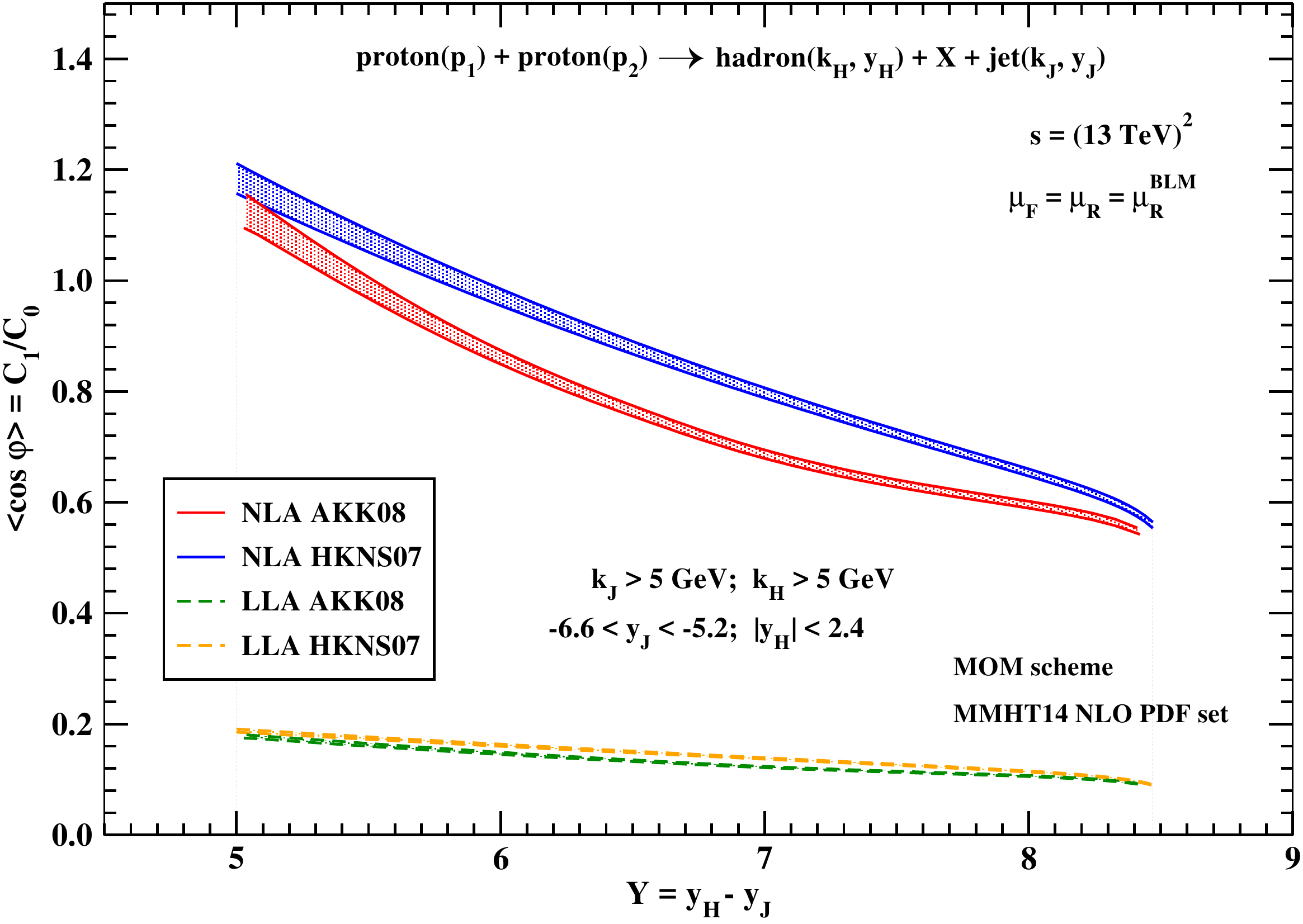}
  
  \includegraphics[scale=0.26,clip]{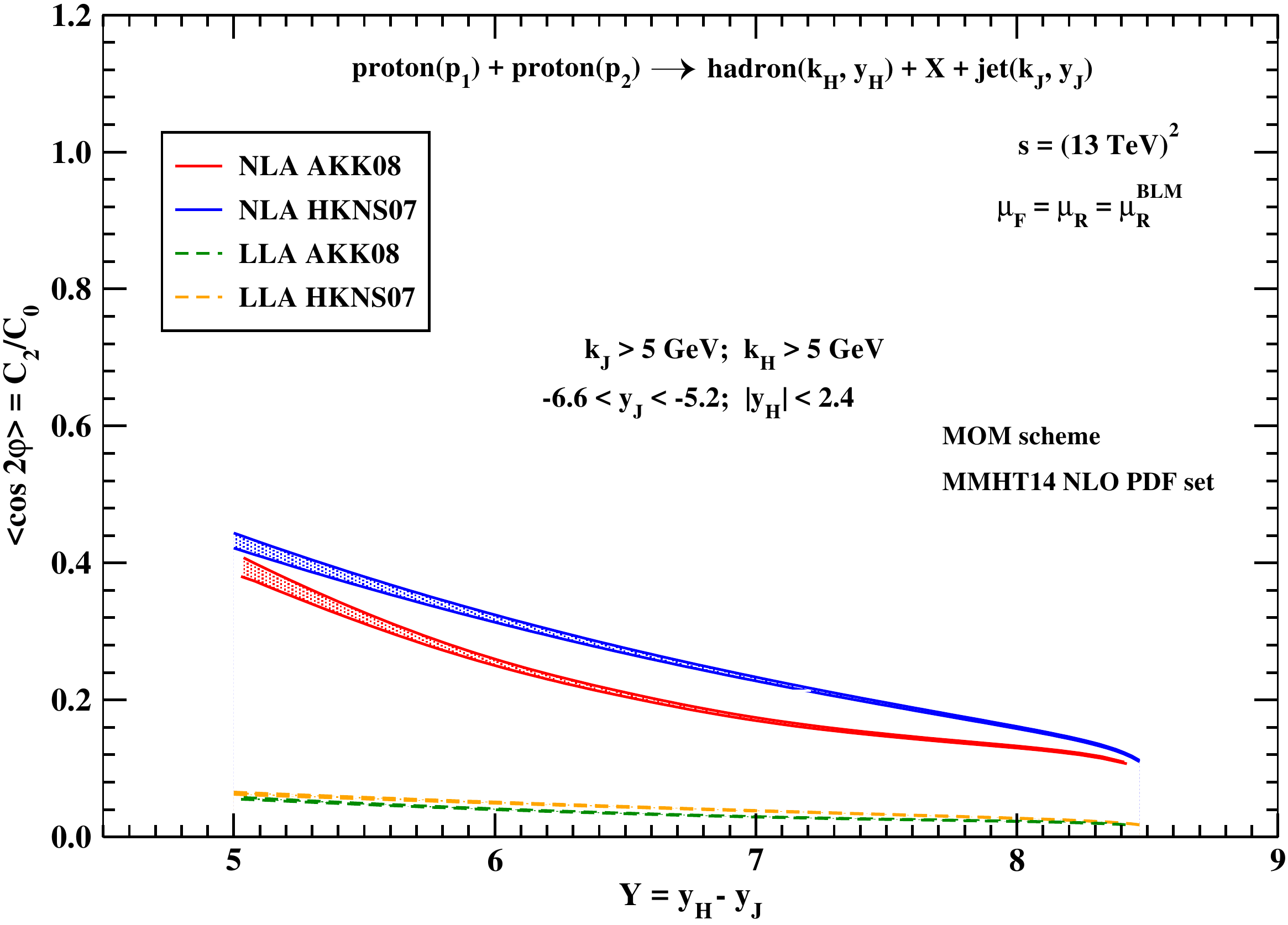}
  \includegraphics[scale=0.26,clip]{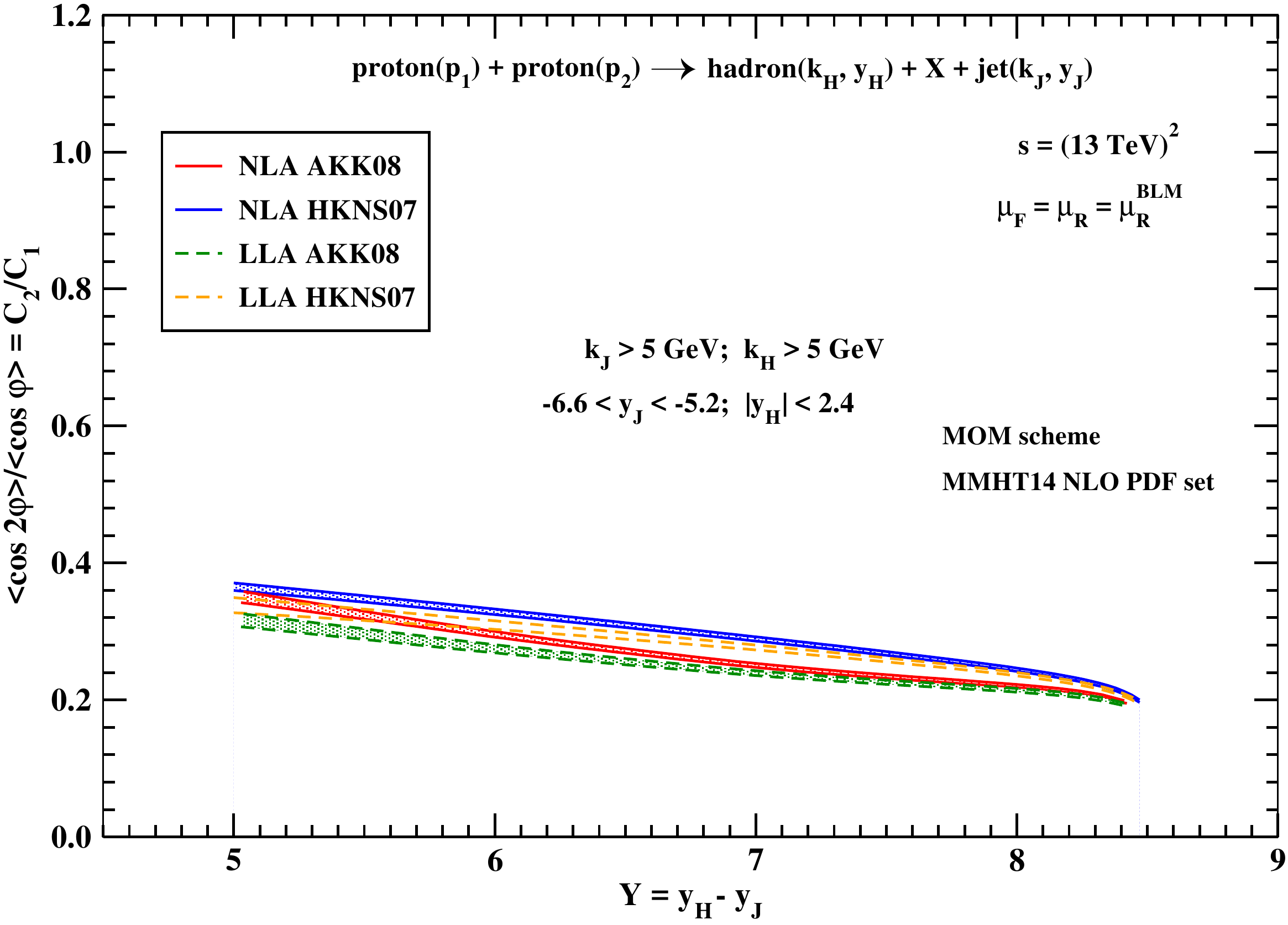}

  {\it b}) {\it CASTOR-jet} configuration.
  \vspace{0.25cm}

  \caption{$Y$-dependence of $C_0$ and of several ratios $C_m/C_n$ for $\sqrt{s} = 13$ TeV.}

  \label{fig:Cn_MOM_BLM_13}
\end{figure}

\begin{figure}[t]
  \centering
  \includegraphics[scale=0.26,clip]{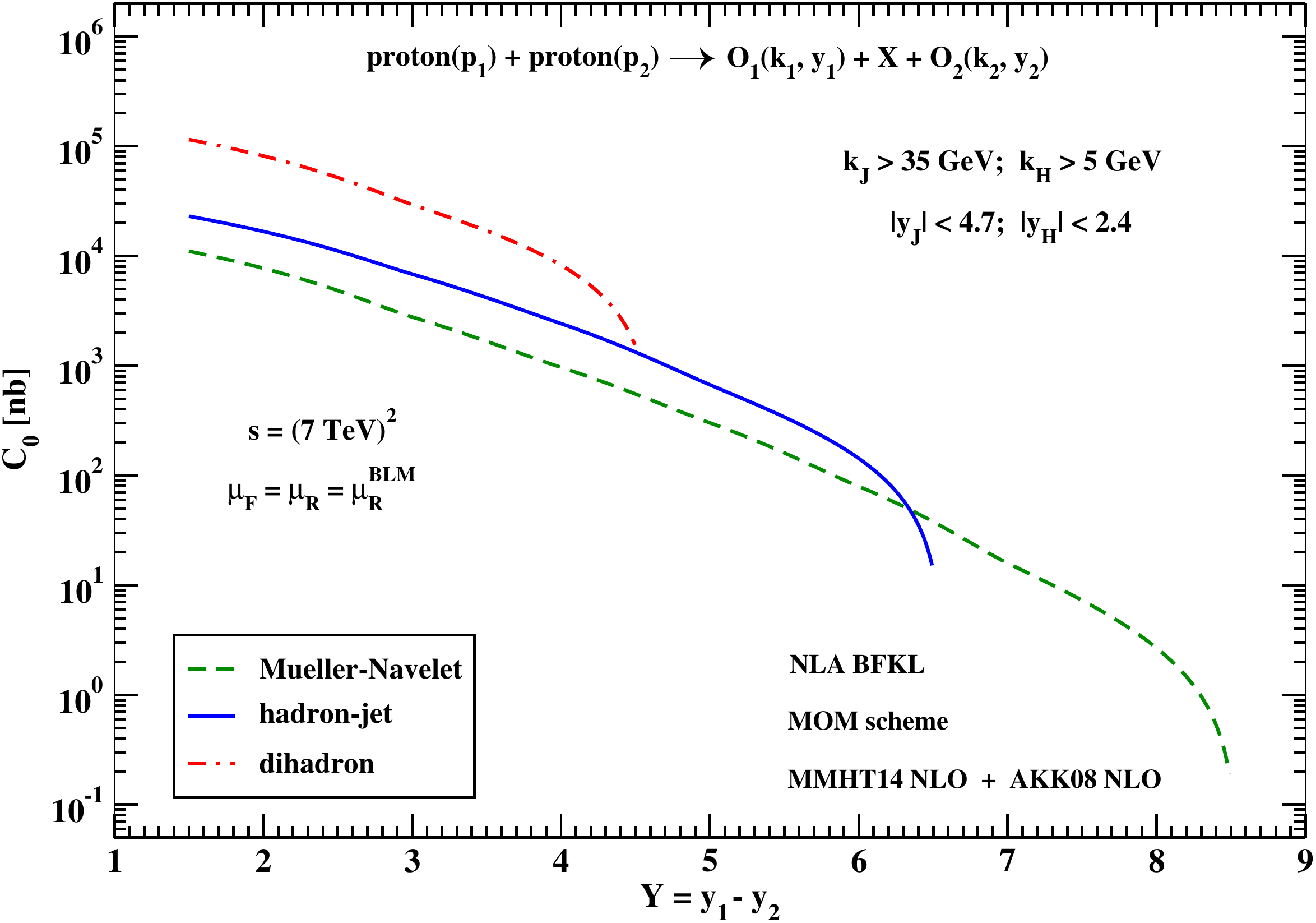}
  \includegraphics[scale=0.26,clip]{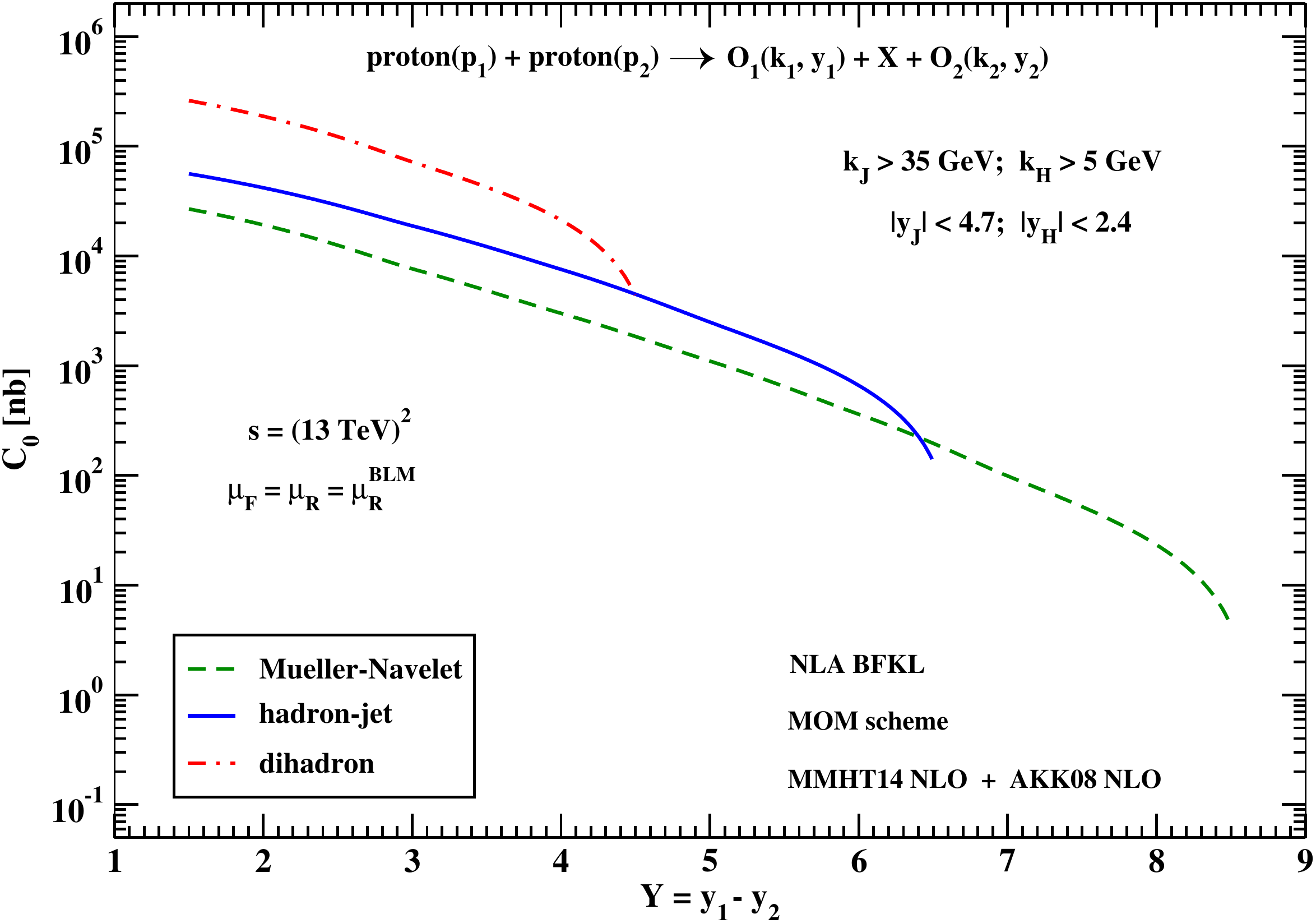}	
  \caption{Comparison of the $\phi$-averaged cross section $C_0$ in different NLA BFKL processes for $\sqrt{s} = 7$, 13 TeV in the {\it CMS-jet} configuration.}
  \label{fig:C0_comp_NLA_BLM_CMS}
\end{figure}

We consider two distinct final-state ranges:
\begin{itemize}
 \item[{\it a})]
  \textit{\textbf{CMS-jet}}~\cite{Khachatryan:2016udy}: 
  both the hadron and the jet tagged by the CMS detector in their typical
  kinematic configurations, {\it i.e.}:
  5 GeV $ < k_H < $ 21.5 GeV, 35 GeV $ < k_J < $ 60 GeV, %\\
  $|y_H| \leq 2.4$, %\\
  $|y_H| \leq 4.7$;
 \item[{\it b})]
  \textit{\textbf{CASTOR-jet}}~\cite{CMS:2016ndp}: a hadron always detected inside CMS in the range given above, together with a very backward jet tagged by CASTOR, lying in the range 5 GeV $ < k_J \lesssim $ 17.68 GeV, %\\
  $-6.6 < y_J < -5.2$.
\end{itemize}
We made all calculations with {\tt JETHAD}, a  \textsc{Fortran} code we recently developed, suited for the study of semi-hard reactions. We used the {\tt MMHT}~2014 NLO PDF set~\cite{Harland-Lang:2014zoa} and two different NLO hadron FFs: {\tt AKK}~2008~\cite{Albino:2008fy} and {\tt HKNS}~2007~\cite{Hirai:2007cx}. We took $\mu_F$ = $\mu_R$ and used the Brodsky-Lepage-Mackenzie (BLM) scheme~\cite{Brodsky:1996sgBrodsky:1997sdBrodsky:1998knBrodsky:2002ka} as derived in its ``exact'' version in~\cite{Caporale:2015uva}. All calculations were done in the MOM renormalization scheme. Predictions for $C_0$ and for several ratios $R_{nm} \equiv C_n/C_m$ at $\sqrt{s} =$ 13 TeV are shown in Fig.~\ref{fig:Cn_MOM_BLM_13}{\it a}) (\textit{CMS-jet}) and Fig.~\ref{fig:Cn_MOM_BLM_13}{\it b}) (\textit{CASTOR-jet}), while in Fig.~\ref{fig:C0_comp_NLA_BLM_CMS} we compare $C_0$ in different NLA BFKL processes: Mueller-Navelet jet, hadron-jet and dihadron production at $\sqrt{s} = 7$, 13 TeV in the {\it CMS-jet} case.  We refer to Sections~3.3 and~3.4 of~\cite{Bolognino:2018oth} for a detailed discussion of results, numerical tools and estimation of the uncertainties.

\section{Summary}

A new candidate probe of the BFKL mechanism, {\it i.e.} the inclusive hadron-jet production at the LHC, has been studied in the NLA accuracy. Distributions over the final-state rapidity interval are in accordance with the ones found for previously investigated semi-hard reactions, when the jet is detected by CMS, while new and interesting aspects emerged when the jet is tagged by CASTOR, which demand further, dedicated analyses.

\section*{Acknowledgments}

F.G.C. acknowledges partial support from the Italian Foundation ``Angelo~della~Riccia'' and from the Italian Ministry of Education, Universities and Research under the FARE grant ``3DGLUE'' (n. R16XKPHL3N).
\\
D.I. thanks the Dipartimento di Fisica dell'U\-ni\-ver\-si\-t\`a della Calabria
and the Istituto Nazio\-na\-le di Fisica Nucleare (INFN), Gruppo collegato di
Cosenza, for the warm hospitality and the financial support.

%uncomment the following lines to place a figure
%\begin{figure}[htb]
%\centerline{%
%\includegraphics[width=12.5cm]{Fig1}}
%\caption{Plot of ...}
%\label{Fig:F2H}
%\end{figure}

\end{document}